\documentclass[aps,nofootinbib,preprintnumbers,prd,superscriptaddress]{revtex4}

\usepackage{mathrsfs}
\usepackage{graphicx}
\usepackage{float}
\usepackage{bm}
\usepackage{hyperref}
\usepackage{amsmath}
\usepackage{amssymb}
\usepackage{lipsum}
\usepackage{mathtools}

\newcommand{\Pslash}{\rlap{$\hspace{.34ex}/$}P}
\newcommand{\kslash}{\rlap{$\hspace{.06ex}/$}k}
\newcommand{\Dslash}{\rlap{$\hspace{.38ex}/$}\Delta}

\begin{document}

\title{Quasi parton distribution function and quasi generalized parton distribution of the pion meson in a spectator model}
\author{Zhi-Lei Ma}
\affiliation{Department of Physics, Yunnan University, Kunming 650091, China}
\author{Jia-qing Zhu}
\affiliation{Department of Physics, Yunnan University, Kunming 650091, China}
\author{Zhun Lu}
\email{zhunlu@seu.edu.cn}
\affiliation{School of Physics, Southeast University, Nanjing 211189, China}

\begin{abstract}
We study the leading-twist quasi parton distribution function (quasi-PDF) and quasi generalize parton distribution (quasi-GPD) of the pion meson by using a spectator model.
We consider the case the quasi functions are defined via inserting the matrix $\gamma_z$ in the spacial correlation functions.
We obtain the analytical expressions for the quasi-PDF and quasi-GPD.
The numerical results for them are calculated from the parameters obtained by fitting the model results to the known parametrization.
Particularly, we investigate the hadron momentum dependence of the quasi functions as well as compare the quasi distributions to the standard functions at different hadron momentum.
We find that in the region $x>0.2$, the quasi distribution are similar to the standard distributions in size and shape when the hadron momentum is larger than $2$ GeV.
Our study thus supports the idea of using quasi distributions to obtain standard distributions.
\end{abstract}

\maketitle

\section{Introduction}
\label{Sec.intro}

The concept of quasi-PDFs for hadrons has been proposed recently in the seminal papers by Ji~\cite{Ji:2013dva, Ji:2014gla} and has received a lot of attention.
Different from the standard parton distribution functions (PDFs) which are defined through the light-cone correlation function, quasi-PDFs are defined by the bilocal operator on a spatial interval such that they can be calculated by the lattice QCD in a four-dimensional Euclidian space.
Standard PDFs, which are of fundamental importance in haronic physics, describe the density that a parton carries in hadron a light-cone fraction $x$ of the total momentum, while quasi-PDFs have similar interpretation, but for a parton carrying a fraction $x$ of the finite momentum $\vec P$ of the hadron.
Although introducing quasi-PDFs will bring a explicit dependence on the hadron momentum (usually denoted by $P^z = |\vec P|$),
it is found that in large limit of $P^z$, the quasi-PDF converges to the standard PDFs.
This provides a convenient way to calculate the $x$-dependence of the standard PDFs by using Lattice QCD.
In particular, a number of lattice calculations on quasi-PDFs and related quantities~\cite{Lin:2014zya, Alexandrou:2015rja, Alexandrou:2016jqi, Chen:2016utp, Zhang:2017bzy, Alexandrou:2017huk, Chen:2017mzz, Green:2017xeu, Lin:2017ani, Orginos:2017kos, Bali:2017gfr, Alexandrou:2017dzj, Chen:2017gck, Alexandrou:2018pbm, Chen:2018xof, Alexandrou:2018eet, Liu:2018uuj, Bali:2018spj, Lin:2018qky}. have been performed.
Meanwhile, to explore for what values of $P^z$ the quasi-PDFs and the standard PDFs are approximation of each other, several model calculations of quasi-PDFs have been carried out~\cite{Gamberg:2014zwa, Bacchetta:2016zjm, Nam:2017gzm, Broniowski:2017wbr, Hobbs:2017xtq, Broniowski:2017gfp, Xu:2018eii,Son:2019ghf}.

Recently, the concept of quasi-PDFs has also been extended to the case of generalize parton distributions (GPDs)~\cite{Ji:1996ek, Radyushkin:1996nd, Ji:1996nm}.
It was proposed in parallel with the quasi-PDFs in Ref.~\cite{Ji:2013dva}.
GPDs are powerful tool developed to characterize the hadron structure two decays ago years (For reviews, see Refs.~\cite{Diehl:2003ny, Belitsky:2005qn, Goeke:2001tz, Guidal:2013rya, Boffi:2007yc, Mueller:2014hsa}).
They unify the conventional notions of parton densities, form factors and distribution amplitudes, which are widely applied in the study of hadron structure.
They also enter the descriptions of various hard exclusive processes~\cite{Ji:1996ek, Radyushkin:1996nd, Ji:1996nm, Radyushkin:1996ru, Collins:1996fb, Collins:1998be}, and they are key objects toward a precision three-dimensional imaging~\cite{Burkardt:2000za,Ralston:2001xs,Diehl:2002he,Burkardt:2002hr} of partons inside the hadron.
Similar to quasi-PDFs, quasi-GPDs may be also calculated on a four-dimensional lattice within QCD.
Although there is no direct lattice calculation of qausi-GPD of the nucleon, the matching procedure between the quasi-GPD and the standard GPD has been developed~\cite{Ji:2015qla, Xiong:2015nua,Liu:2019urm} by using large momentum effective theory.
On the other hand, quasi-GPDs of the nucleon has been studies by models~\cite{Bhattacharya:2018zxi, Bhattacharya:2019cme}, which may provide useful theoretical constraints on those objects.

In this paper, we will investigate both the quasi-PDFs and the quasi-GPDs of the pion meson using a spectator-antiquark model.
Although the pion meson is the lightest hadron, it still remains as a challenge to determine its nonperturbative properties, including its partonic structure~\cite{Holt:2010vj,Aguilar:2019teb}.
Compared to nucleons, the structure of the pion meson has not been probed extensively by experiments due to the fact it can not been served as a target in high energy scattering processes.
It thus largely relies on theoretical studies as well as lattice calculations which are important for obtaining the information on the pion structure.
As the standard distributions of the pion are the mapping of the quasi distributions in large $P^z$ limit, the study of quasi-distributions paves the way to gain information of the pion structure through lattice calculations.
Particularly, the first lattice results for the quasi-PDF and the quasi-GPD (at zero-skewness) of the pion meson are available~\cite{Chen:2018fwa,Chen:2019lcm}.
In the present work, we will focus on the leading-twist chiral-even quasi-PDF and quasi-GPD of the pion meson.
The higher-twist or chiral-odd quasi-distributions can be studied in a similar way.
The main purpose of our study is to understand, to what extent, the quasi-PDFs and the quasi-GPDs of the pion can approximately describe by each other from the viewpoint of models.

We organize the paper as follows:
In Sec.~II, we set up the kinematics and review the operator definitions of quasi-PDFs and quasi-GPDs.
In Sec.~III, we analytically calculate the quasi-PDF $\tilde{f}_{1\pi}(x, P^z)$ and the quasi-GPD $F_{1\pi}(x,\xi,\Delta_\perp, P^z)$ of the pion meson in the spectator-antiquark model.
The results of the standard PDF and the standard GPD are also given for later comparison.
In Sec.~IV, we provide the numerical results for the quasi-PDF and the quasi-GPD and explore the $x$ and $P^z$ dependence of the quasi distributions.
The skewness dependence of the quasi-GPD is also presented.
Finally, in Sec.~V we summarize our work.

\section{Kinematics and definitions}

In this section, we shortly review the standard PDF as well as the GPD of the pion meson at leading-twist.
The Standard PDF is usually defined as the bilocal operator on the light-cone:
\begin{align}
f_1(x) = \int {d \zeta^- \over 4\pi} e^{-x\zeta^- P^+} \langle P | \bar{\psi}(\zeta^-) \gamma^+
\mathcal{L}[\zeta^-, 0] \psi(0)|P\rangle. \label{eq:pdf-def}
\end{align}
with $P$ denoting the four-momentum of the pion moving along the $z$-axis with the components $(P^+, P^-, \bm 0_\perp)$ in the light-cone coordinates, in which the plus and minus components of any four-vector $a^\mu$ have the form $a^{\pm} =(a^0 \pm a^3)/\sqrt{2}$, and transverse part $\bm a_\perp =(a^1, a^2)$.

In Eq.~(\ref{eq:pdf-def}), $\mathcal{L}[\zeta^-, 0]$ is the gauge-link insuring the gauge-invariance of the operator definition:
\begin{align}
\mathcal{L}[\zeta^-, 0]  = \mathcal{P}\exp\left(-ig\int_{0}^{\zeta^-} d\eta^-  A^+(\eta^-)\right),\label{eq:pdfgauge}
\end{align}
where $\mathcal{P}$ denotes the path ordering of the gauge-link, $A^+$ the plus-component of the gluon field, $g$ the coupling constant.

Similarly, standard GPDs can be defined through the following light-cone quark-quark correlator
\begin{align}
F^{[\Gamma]}(x,\Delta) =& \int {d \zeta^-\over 4\pi} e^{-ik^+ \zeta^-}
\langle p^\prime | \bar{\psi}(\tfrac{\zeta}{2}) \,\Gamma\,
\mathcal{L}[\tfrac{\zeta^-}{ 2}, -\tfrac{\zeta^-}{2}] \psi(-\tfrac{\zeta}{2}) | p\rangle\, ,  \label{eq:gpd-corr}
\end{align}
where $\Gamma$ denotes a generic gamma matrix, $\Delta = p^\prime - p = (\Delta^+, \Delta^-, \bm
\Delta_\perp)$ is the momentum transfer of the hadron.
The color gauge invariance of the correlator is again ensured by the gauge-link $\mathcal{L}[\tfrac{\zeta^-}{2}, -\tfrac{\zeta^-}{2}]$ similar to (\ref{eq:pdfgauge}), but running from $-\tfrac{\zeta^-}{2}$ to $\tfrac{\zeta^-}{2}$ along the minus light-cone component:
\begin{align}
\mathcal{L}[\tfrac{\zeta^-}{2}, -\tfrac{\zeta^-}{2}]  = \mathcal{P}\exp\left(-ig\int_{\tfrac{\zeta^-}{2}}^{-\tfrac{\zeta^-}{2}} d\eta^-  A^+(\eta^-)\right).\label{eq:pdfgauge}
\end{align}
The four-momenta of the initial-state and final-state hadrons are represented by $p$ and $p^\prime$, respectively.

For GPDs one usually works in the so-called symmetric frame in which the commonly used notations for the kinematical variables are
\begin{align}
P={1\over 2}(p+p^\prime) =(P^+, P^-, \bm 0_\perp), ~~~~\xi = {p^{\prime+} -p^+\over p^\prime +p^+} = - {\Delta^+\over 2 P^+},~~~~t= \Delta^2 = -{1\over 1-\xi^2} \left(4\xi^2 M^2 + \bm \Delta_\perp^2\right).
\end{align}
where $\xi$ is the skewness variable in the range $0\le \xi \le 1$, $M$ is the mass of the hadron.

For spin-0 hadron such as the pion meson, in twist-2 there is one chiral-even GPD which can be defined by setting
 $\Gamma = \gamma^+$ in Eq.~(\ref{eq:gpd-corr}):
\begin{align}
F_1^\pi (x,\xi,\Delta_\perp) =& \int {d \zeta^-\over 4\pi} e^{-ik^+ \zeta^-}
\langle p^\prime | \bar{\psi}\left(\tfrac{\zeta^-}{2}\right) \,\gamma^+\,
\mathcal{L}\left[\tfrac{\zeta^-}{ 2}, -\tfrac{\zeta^-}{ 2}\right]
\psi\left(\tfrac{-\zeta}{2}\right) |p\rangle\,,
\end{align}
where the skewness variable separates the $x$ region $-1 \le x \le 1$ to two catalogs, one is the DGLAP region for quarks (antiquarks) $\xi \le x \le 1$ ($-1 \le x \le -\xi$), the other is the ERBL region $-\xi \le x \le \xi$.

In the following we turn to the definitions of the quasi-PDF and quasi-GPD of the pion meson.
The quasi-PDFs are equal-time spatial correlations along the $z$-direction, and the corresponding operator definition has the following form
\begin{align}
f_1(x;P^z) = \int {d \zeta_z \over 4\pi} e^{-i x\zeta_z P^z} \langle P | \bar{\psi}(\zeta_z) \gamma_z
\mathcal{L}_{n_z}[\zeta_z, 0] \psi(0)|P\rangle. \label{eq:qpdf-def}
\end{align}
where we used the same notation for the fractional momentum of the quark inside a hadron moving along the $z$ direction:  $x = {k^z\over P^z}$.
For quasi-PDFs the Minkowski components are used for any four-vector: $a^\mu=(a^0, \bm a_\perp, a^z)$, in which the hadron momentum is $P=(P^0, \bm 0_\perp, P^z)$.
In addition, $\mathcal{L}_{n_z}$ in Eq.~\ref{eq:qpdf-def} is the gauge-link insuring the gauge-invariance of the operator definition:
\begin{align}
\mathcal{L}_{n_z}[\zeta_z, 0]  = \exp\left(\int_{\zeta_z}^{0} d\eta_z  A_z(\eta_z)\right)
\end{align}
where $n_z = (0,\bm{0}_\perp, 1)$ and $n_z \cdot v$ for any four-vector $v_\mu$

Finally, the twist-2 chiral-even quasi-GPD of the pion is defined through the following spatial correlator:
\begin{align}
F_{1\pi} (x,\xi,t,P^z) =& \int {d \zeta_z\over 4\pi} e^{-ik^z \zeta_z}
\langle p^\prime | \bar{\psi}\left(\tfrac{\zeta_z}{2}\right) \,\gamma_z\,
\mathcal{L}\left[\tfrac{\zeta_z}{2}, -\tfrac{\zeta_z}{2}\right] \psi\left(-\tfrac{\zeta_z}{2}\right) |p\rangle
\end{align}

\section{Analytic calculations}

In this section, we apply the spectator model to perform the analytic calculation on the quasi-PDF $f_{1\pi}(x;P^z)$ and quasi-GPD $F_{1\pi} (x,\xi,\Delta_\perp;P^z)$ of the pion.
The spectator model~\cite{Jakob:1997wg} has been widely applied to study the (TMD)PDFs~\cite{Boer:2002ju, Gamberg:2003ey, Bacchetta:2003rz, Lu:2004au, Gamberg:2007wm, Bacchetta:2008af} and GPDs~\cite{Meissner:2007rx} of the nucleon as well as those of the spin-0 hadron~\cite{Lu:2004hu, Meissner:2008ay, Gamberg:2009uk}.
The spectator model has also been used to calculate the unpolarized/polarized quasi-PDFs~\cite{Gamberg:2014zwa} and quasi-GPDs~\cite{Bhattacharya:2018zxi, Bhattacharya:2019cme} of the nucleon recently.
In the spectator model, the Fock-state of the hadron can be truncated as the active quark and an intermediate spectator.,
In the case of the pion, the spectator is served by the antiquark, and the coupling between the pion and the quark-antiquark pair is characterized by a pseudo-scalar interaction.
Thus, the interaction part of the Lagrangian can be written as (including the isospin)
\begin{align}
 \mathcal{L}_\text{int}(x) & = - i \, g_{\pi} \, \bar{\Psi}(x) \, \gamma_5 \,
 \bm{\tau} \cdot \bm{\varphi}(x) \, \Psi(x) \,,
\end{align}
where $g_\pi$ is the coupling of the pion-quark-antiquark vertex.
To smoothly suppress the influence of large $k_\perp$ and eliminate the divergences arising after
integration over $k_\perp$ when using a point-like coupling, we replace $g_\pi$ with a exponential form factor as follows
\begin{align}
g_\pi \mapsto g_\pi(k_\perp)=g_\pi^\prime \exp\left(-{k_\perp^2 \over \bar{x}^\alpha (1-\bar{x})^\beta\lambda^2}\right) \equiv
g_\pi^\prime \exp\left(-{k_\perp^2 \over \Lambda^2(x)}\right), \label{eq:expform}
\end{align}
where $\bar{x}=|x|$, and $\Lambda(x)=\bar{x}^\alpha (1-\bar{x})\beta\lambda$ can be understood as the cutoff parameter, with $\alpha$ and $\beta$ the parameters of the model.
We note that in Ref.~\cite{Gamberg:2014zwa} the dipolar form factor was adopted to calculate the quasi-PDFs of the nucleon in the spectator model, while in Ref.~\cite{Bhattacharya:2018zxi}, a cutoff $\Lambda= 1$ GeV was chosen for the $k_\perp$ integration.
By choosing the form factor in Eq.~\ref{eq:expform}, we have limited the applicable range of our model to the region $-1\leq x\leq 1$, since this is the most interesting kinematical region for quasi-PDFs and quasi-GPDs.
In Ref.~\cite{Bhattacharya:2018zxi}, the calculation of quasi-GPDs has been extended to the region $|x|>1$.

\subsection{Twist-2 standard PDF and GPD of the pion meson}

For the unpolarized PDF of the pion meson, the corresponding expression for the calculation is
\begin{align}
& f_{1\pi}(x) =  \, \int  { dk^- d^2\bm k_\perp \over 2 (2\pi)^4 }
g_{\pi}^2(\bm k_\perp^2) { \text{Tr} \big[ (\Pslash - \kslash + M_q) \,
          (\kslash  + M_q) \,
          \gamma^+ \, (\kslash + M_q) \big]
          \over ((P-k)^2-M_q^2+i\epsilon)(k^2-M_q^2+i\epsilon)^2} \,,\label{eq:corr_pdf}
\end{align}
where $M_q$ is the quark/antiquark mass. The integration over $k^-$ can be carried out by using the replacement:
\begin{equation}
{1\over (P-k)^2-m_q^2+i\epsilon)} \rightarrow  {-2\pi i\over 2(1-x)P^+ }\delta\left(k^- -{M_\pi^2 \over 2P^+} + {k_\perp^2+ m_q^2 \over 2(1-x)P^+} \right)
\end{equation}
which yields
\begin{equation}
f_{1\pi}(x) = g_\pi^{\prime 2}\int {d^2 \bm k_\perp\over (2\pi)^3} { (\bm k_\perp^2 +M_q^2)\over (\bm k_\perp^2 +M_q^2-x(1-x)M^2)^2}  \exp\left(-{2\bm k_\perp^2\over \Lambda^2(x)}\right)
\end{equation}
Next we perform the $\bm k_\perp$-integration analytically and obtain the following result
\begin{equation}
f_{1\pi}(x) = {g_\pi^{\prime 2}\over 8\pi^2} \left[{ x(1-x)M^2\over M^2(x)}-
\left({2x(1-x)M^2\over\Lambda(x)^2}-1\right)
 \exp\left({2M^2(x) \over \Lambda^2(x)}\right)\Gamma\left(0,{2M^2(x) \over \Lambda^2(x)}\right)\right],
 \label{eq:pipdf}
\end{equation}
where $M^2(x)=M_q^2-x(1-x)M^2$,  $\Gamma(0,x)$ is the zeroth-order incomplete Gamma function.

On the other hand, the twist-2 chiral-even GPD for a spin-0 hadron, denoted by $F_{1\pi}(x, \xi, \Delta_\perp)$~\cite{Meissner:2008ay}, can be calculated from the following correlator
\begin{align}
&F_{1\pi}(x,\xi,\bm \Delta_\perp)=
\int { d k^- d^2\bm k_\perp\over 4(2\pi)^3} { g_\pi^+g_\pi^-\text{Tr} \big[ (\Pslash - \kslash + m) \, (\kslash + \tfrac{1}{2} \Dslash + M_q) \,
\gamma^+ \, (\kslash - \tfrac{1}{2} \Dslash + m) \big] \over [(P-k)^2-m^2+i\epsilon][(k-\tfrac{1}{2}\Delta)^2-M_q^2+i\epsilon]
[(k+\tfrac{1}{2}\Delta)^2-M_q^2+i\epsilon] } \,,\label{eq:corr_gpd}
\end{align}
where $g_\pi^{\pm}= g_\pi\left(k_\perp\pm \tfrac{1}{2}\Delta_\perp\right)$
When calculating GPDs, apart from the cut through the spectator, those through the active quarks should also be included~\cite{Meissner:2009ww}.
After some algebra, the results for $F_{1\pi}(x, \xi, \Delta_\perp)$ can be organized as~\cite{Meissner:2008ay}
\begin{equation}
F_{1\pi}(x, \xi, \Delta_\perp) =
\begin{dcases}
0 & \quad -1 \le x \le -\xi  \,, \\[0.15cm]
\frac{g_\pi^{\prime 2} (x + \xi) (1 + \xi) (1 - \xi^2)}{2(2 \pi)^3 (1-x)}\int d^2\bm{k}_\perp \, \frac{N(x,\xi,\bm\Delta_\perp,\bm k_\perp)}{D_1 \, D_2^{- \xi \le x \le \xi}}
  \, \exp\left(-{2\bm k_\perp^2+\tfrac{1}{2}\bm \Delta_\perp^2\over \Lambda^2(x)}\right)& \quad -\xi \le x \le \xi  \,, \\[0.15cm]
\frac{g_\pi^{\prime 2}  (1 - \xi^2)}{ (2 \pi)^3} \int d^2\bm{k}_\perp \, \frac{N(x,\xi,\bm\Delta_\perp,\bm k_\perp)}{D_1 \, D_2^{x \ge \xi}}\exp\left(-{2\bm k_\perp^2+\tfrac{1}{2}\bm \Delta_\perp^2\over \Lambda^2(x)}\right)  & \quad x \ge \xi \,,
\end{dcases} \label{eq:pigpd}
\end{equation}
where the numerator $N(x,\xi,\bm\Delta_\perp,\bm k_\perp)$ has the form:
\begin{align}
N(x,\xi,\bm\Delta_\perp,\bm k_\perp)=
(1-\xi^2)(\bm k_\perp^2+M_q^2)+(1-x)\xi \bm k_\perp \cdot\bm \Delta_\perp-((1-x)^2)\frac{1}{4}\bm \Delta_\perp^2
\end{align}
and the $D_i$ in the denominator
\begin{eqnarray}
D_1 & = & (1 + \xi)^2 \bm k_\perp^2 + \frac{1}{4} (1 - x)^2 \bm{\Delta}_\perp^2 - (1 - x) (1 + \xi) \bm{k}_\perp \cdot \bm{\Delta}_\perp + (1 + \xi)^2 M_q^2
- \, (1 - x) (x + \xi) M^2 \,, {\phantom{\frac{1}{4}}}
\nonumber \\
D_2^{- \xi \le x \le \xi} & = & \xi (1 - \xi^2) \bm k_\perp^2 + \frac{1}{4} (1 - x^2) \xi \bm{\Delta}_\perp^2 + x (1 - \xi^2) \bm{k}_\perp \cdot \bm{\Delta}_\perp + \xi (1 - \xi^2) M_q^2 - \xi (x^2 - \xi^2) M^2 \,,
\nonumber \\
D_2^{x \ge \xi} & = & (1 - \xi)^2 \bm k_\perp^2 + \frac{1}{4} (1 - x)^2 \bm{\Delta}_\perp^2 + (1 - x) (1 - \xi) \bm{k}_\perp \cdot \bm{\Delta}_\perp +  (1 - \xi)^2 M_q^2 - \, (1 - x) (x - \xi) M^2 \,. {\phantom{\frac{1}{4}}}
\end{eqnarray}
where $D_2^{x \ge \xi}$ comes from the cut through the spectator antiquark, and $D_2^{- \xi \le x \le \xi} $ corresponds to the cut through the quark line, which gives the contributions in the ERBL region.

\subsection{twist-2 Quasi-PDF and quasi-GPD of the pion meson}

According to Eq.~\ref{eq:qpdf-def}, the quasi-PDF of the pion meson in the spectator model can be calculated from
\begin{equation}
 f_{1\pi}(x;P^z) = - \, \int  { dk^0 d^2\bm k_\perp \over 2(2\pi)^4 }g_{\pi}^{2}(k_\perp)
{ \text{Tr} \big[ (\Pslash - \kslash + M_q) \,
          (\kslash  + M_q) \,
          \gamma_z \, (\kslash + M_q) \big]
          \over ((P-k)^2-M_q^2+i\epsilon)(k^2-M_q^2+i\epsilon)^2} \,,\label{eq:corr_pdf}
\end{equation}

Analyzing the pole structure of the integrand in Eq.~\ref{eq:corr_pdf}, we can rewrite the denominator as
\begin{align}
{1\over((P-k)^2-M_q^2+i\epsilon)(k^2-M_q^2+i\epsilon)^2} =
{1\over(k^0 - k_-^0)^2 (k^0 - k_+^0 )^2(k^0 - k_-^{\prime 0} ) (k^0 - k_+^{\prime 0})}
\end{align}
where $k_{\pm}^0$ are the poles for $k^0$:
\begin{align}
k_{\pm}^0 &= \pm \sqrt{x^2 (P^z)^2+\bm k_\perp^2+M_q^2-i\epsilon}\\
k_{\pm}^{\prime 0} &= \delta_0 P^z \pm \sqrt{(1-x)^2 (P^z)^2+ \bm k_\perp^2+M_q^2-i\epsilon},
\end{align}
with
\begin{equation}
\delta_0= P^0/P^z = \sqrt{1 + {k_\perp^2 + M_\pi^2\over P_z^2}} ,
\end{equation}

Performing the integration over $k^0$ using the residue theorem, on can obtain $f_{1\pi}(x;P^z)$ as
\begin{eqnarray}
f_{1,\pi}(x; P^z) & = & - \, \frac{1}{(2\pi)^3} \int d^2 \bm{k}_\perp g_\pi^2(k_\perp) \bigg[
\frac{N(k_-^{\prime 0})}{(k_-^{\prime 0} - k_+^0)^2 \, (k_-^{\prime 0} - k_-^0)^2 \, (k_-^{\prime 0} - k_+^{\prime 0})}
\nonumber \\[0.1cm]
& & + \, \frac{N^\prime(k_-^0)}{(k_-^0 - k_+^0)^2 \, (k_-^0 - k_+^{\prime 0}) \, (k_-^0 - k_-^{\prime 0})}
- \frac{2 \, N(k_-^0)}{(k_-^0 - k_+^0)^3 \, (k_-^0 - k_+^{\prime 0}) \, (k_-^0 - k_-^{\prime 0})}
\nonumber \\[0.1cm]
& & - \, \frac{N(k_-^0)}{(k_-^0 - k_+^0)^2 \, (k_-^0 - k_+^{\prime 0})^2 \, (k_-^0 - k_-^{\prime 0})}
- \frac{N(k_-^0)}{(k_-^0 - k_+^0)^2 \, (k_-^0 - k_+^{\prime 0}) \, (k_-^0 - k_-^{\prime 0})^2}
\bigg] \,,
\label{eq:quasi-pdf}
\end{eqnarray}
The numerator $N$ in Eq.~\ref{eq:quasi-pdf} has the form
\begin{align}
N(k^0)&=-2(1-x)x^2(P^z)^3+ 4 x\delta k_0 (P^z)^2 -(2(1+x)(k_0^2-k_\perp^2-M_q^2))P^z
\end{align}
and $N^\prime(k^{0})$ is the first-order derivative of $N(k^0)$ coming from the contribution of the double pole
\begin{equation}
 \int_C {N(k_0)\over (k_0-k_{i0})^2  } d k_0= 2\pi i\, {d N(k_0) \over d k_0} \bigg{|}_{k_0=k_{i0}}\,.
\end{equation}
It can be easily verified that in the limit $P^z\rightarrow \infty$, the quasi-PDF in Eq.~\ref{eq:quasi-pdf} reduces to the standard PDF presented in Eq.~\ref{eq:pipdf}.

On the other hand, in literature~\cite{Gamberg:2014zwa,Bacchetta:2016zjm} the cut-diagram approach has also been applied to calculate the quasi-PDF of the proton.
In this framework, a sum over a complete set of states between the quark fields is inserted into the quasi-PDF operator.
Here we apply this approach to calculate the quasi-PDF of the pion for comparison:
\begin{align}
f_{1,\pi}^{\textrm{(cut)}}(x; P^z) & =  - \, \frac{1}{(2\pi)^3} \int d^2 \bm{k}_\perp g_\pi^2(k_\perp) \bigg[
\frac{N(k_-^{\prime 0})}{(k_-^{\prime 0} - k_+^0)^2 \, (k_-^{\prime 0} - k_-^0)^2 \, (k_-^{\prime 0} - k_+^{\prime 0})}\bigg]
\end{align}
This result corresponds to the first term of on the right hand side of Eq.~(\ref{eq:quasi-pdf}).
As shown in Ref.~\cite{Bhattacharya:2018zxi}, this term will only give rise to the quasi-PDF in the positive $x$ region.

The quasi-GPD of the pion meson can be calculated in a similar way.
In the spectator model, the correlator to calculate the quasi-GPD has the form:
\begin{align}
F^{\Gamma} = {i\over 2(2\pi)^4}\int dk^0 d^2\bm k_\perp g_\pi^+ g_\pi^-
{(\Pslash-\kslash+m)(\kslash+{\Dslash\over 2}+m)\Gamma (\kslash-{\Dslash\over 2}+m)\over \left(\left(k+{\Delta\over 2}\right)-m^2+i\epsilon\right)
\left(\left(k+{\Delta\over 2}\right)-m^2+i\epsilon\right)\left((P-k)^2-m^2+i\epsilon\right)}
\end{align}
The quasi-GPD can be define by setting $\Gamma= \gamma_z$ or $\Gamma=\gamma_0$, as shown in Ref.~\cite{Bhattacharya:2018zxi}.
In our calculation we will adopt $\Gamma= \gamma^z$ to calculate the quasi-GPD.
We checked that the two choices are qualitatively agree with each other.

After performing the $k^0$-integral using the contour integration, we write down the analytical result of
the quasi-GPD as follows
\begin{align}
F_{1\pi}(x,\xi,\Delta_\perp;P^z) = {1\over (2\pi)^3}\int d^2 \bm k_\perp g_\pi^+ g_\pi^- \sum_{i=0}^3{N(k_{i-}^0)\over D(k_{i-}^0)}, \label{eq:quasi-gpd}
\end{align}
where the numerator has the expression
\begin{align}
N(k^0)&=2[(1-x)(\xi^2 - x^2) -(1+x)\delta^2 \xi^2](P^z)^3+ 4 (x+\xi^2)\delta k_0 (P^z)^2 \nonumber\\ &-(2(1+x)(k_0^2-k_\perp^2-m_q^2)+(1-x)\Delta_\perp^2/2 - 2\delta \xi k_\perp\cdot\Delta_\perp)P^z,
\end{align}
with $\delta = \sqrt{1+{M_\pi^2-t/4\over P_z^2}}$.
To obtain the above result we have also used the relations
\begin{align}
 \Delta^0 &= - 2 \xi P^z,\;\;  \Delta^3 =- 2 \delta \xi  P^z
\end{align}

\begin{table}[t]
\centering
\begin{tabular}{c|c|c|c}
  \hline
  ~~$g_\pi^\prime $~~& ~~$\lambda$ ~~& ~~$\alpha$~~ & ~~$\beta$~~ \\
  \hline\hline
  $6.316\pm 0.514$ & $0.855\pm0.138$  & 0(fixed) & 1(fixed) \\
  \hline
\end{tabular}
\caption{The values of the parameters in the spectator model fitted to the
GRV parametrization~\cite{Gluck:1991ey}.}
\label{tab:para}
\end{table}

The denominators in (\ref{eq:quasi-gpd}) have the forms
\begin{align}
 D(k_{1-}^0) &= (k_{1-}^0 - k_{1+}^0) (k_{1-}^0 - k_{2+}^0) (k_{1-}^0 - k_{2-}^0) (k_{1-}^0 - k_{3+}^0)
(k_{1-}^0 - k_{3-}^0)\\
 D(k_{2-}^0) &= (k_{2-}^0 - k_{1+}^0) (k_{2-}^0 - k_{1-}^0) (k_{2-}^0 - k_{2+}^0) (k_{2-}^0 - k_{3+}^0)
(k_{2-}^0 - k_{3-}^0)\\
 D(k_{3-}^0) &= (k_{3-}^0 - k_{1+}^0) (k_{3-}^0 - k_{1-}^0) (k_{3-}^0 - k_{2+}^0) (k_{3-}^0 - k_{2-}^0)
(k_{3-}^0 - k_{3+}^0)
\end{align}
The poles for the quark propagators and the antiquark spectator are given by
\begin{align}
k_{1\pm}^0 &= -\xi P^z \pm \sqrt{(x+\delta\xi)^2 (P^z)^2+(\bm k_\perp -{\bm\Delta_\perp \over 2})^2+m^2-i\epsilon}\nonumber\\
k_{2\pm}^0 &= \xi P^z \pm \sqrt{(x-\delta\xi)^2 (P^z)^2+(\bm k_\perp +{\bm\Delta_\perp \over 2})^2+m^2-i\epsilon} \label{eq:polegpd} \\
k_{3\pm}^0 &= \delta P^z \pm \sqrt{(1-x)^2 (P^z)^2+ \bm k_\perp^2+m^2-i\epsilon},\nonumber
\end{align}
To choose the poles $k^0=k_{i-}^0$ means that we close the upper half plane to perform the integration.

\begin{figure}
  \centering
  \includegraphics[width=0.49\columnwidth]{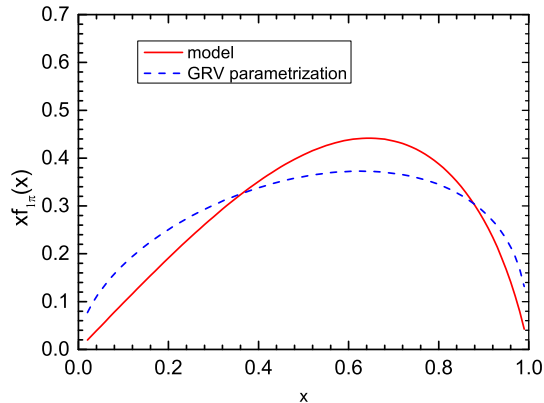}
  \caption{Fit of the pion PDF $f_{1\pi}(x)$ to the GRV parametrization at the scale $Q_0=0.5$ GeV. The solid and the dashed curves depict the model result and the GRV parametrization, respectively.}\label{fig:f1pi}
\end{figure}

The analytical result for the quasi-GPD of the pion meson in the spectator model applies in the whole $x$ region, including the ERBL region $-\xi \le x \le \xi$.
This is different from the spectator model result of the standard GPD, which has different expressions in the DGLAP region and in the ERBL region.
We also verify that in the large limit of $P^z$, the quasi-GPD sketched in Eq.~(\ref{eq:quasi-gpd}) will converge to the standard GPD in Eq.~(\ref{eq:pigpd}).

\section{numerical result}

To obtain a more realistic result for the quasi-PDF and quasi-GPD of the pion meson, we fit our model result of $f_1\pi(x)$ to the known parametrization, e.g., the GRV parametrization~\cite{Gluck:1991ey} for the pion.
As the GRV parametrization does not provide the uncertainties for the PDF, we will take $20\%$ of the PDF value as its uncertainty, which is a reasonable choice.
The fitted values for the parameters together with their uncertainties are given in Table.~\ref{tab:para}.
We adopt the quark mass as $m=0.2$ GeV.
In fig.~\ref{fig:f1pi}, we plot our fit on $f_{1\pi}(x)$ and compared the curve to the
GRV parametrization for $f_{1\pi}$ at the scale $\mu_0= 0.5$ GeV.
The solid line corresponds to the model result of $f_{1\pi}(x)$ calculated from the central values of the parameters. The dashed line denotes the GRV parametrization for $f_{1\pi}(x)$.
A qualitative agreement between the model and the parametrization is found.

\begin{figure*}
  \centering
  \includegraphics[width=0.48\columnwidth]{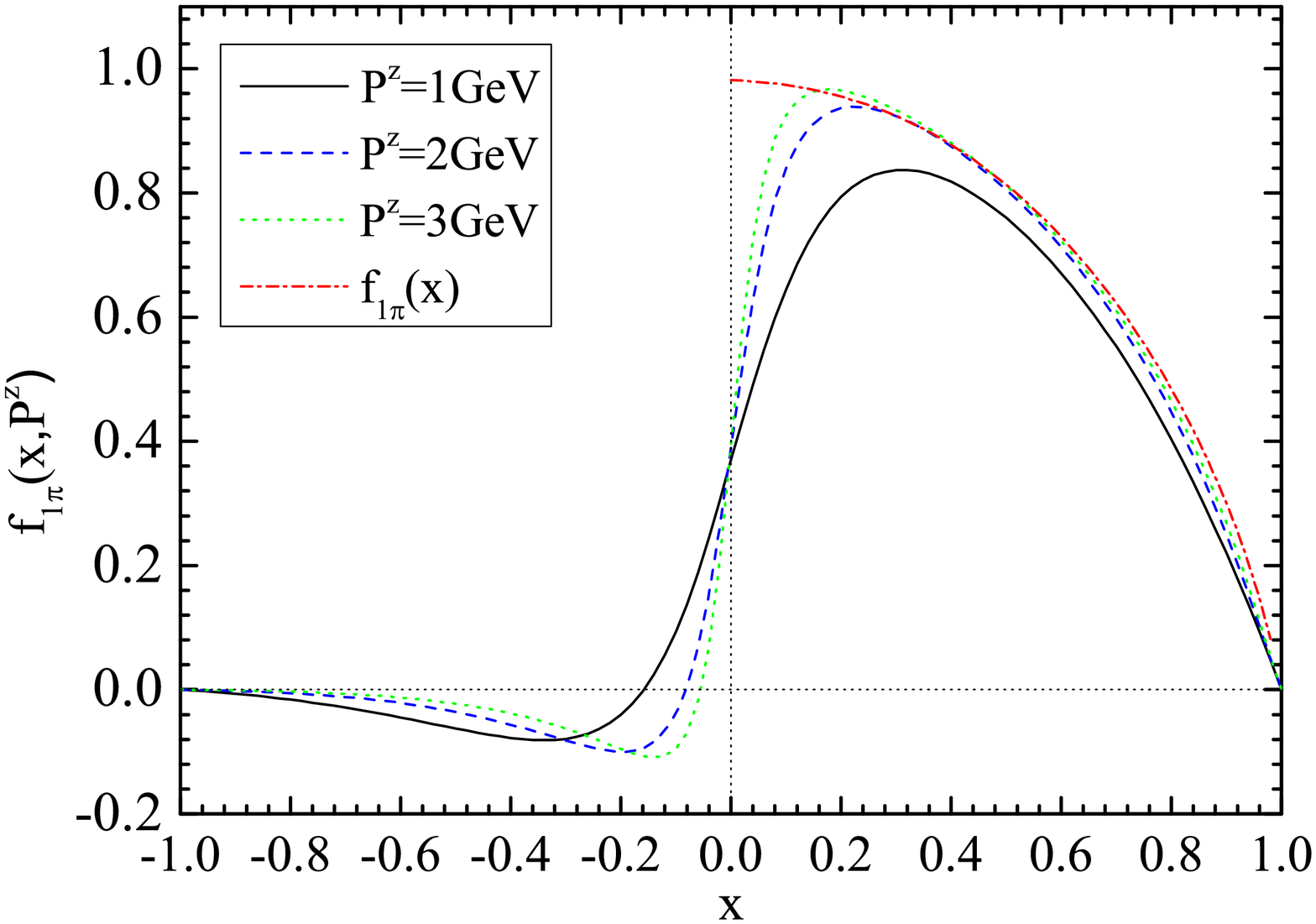}
    \includegraphics[width=0.49\columnwidth]{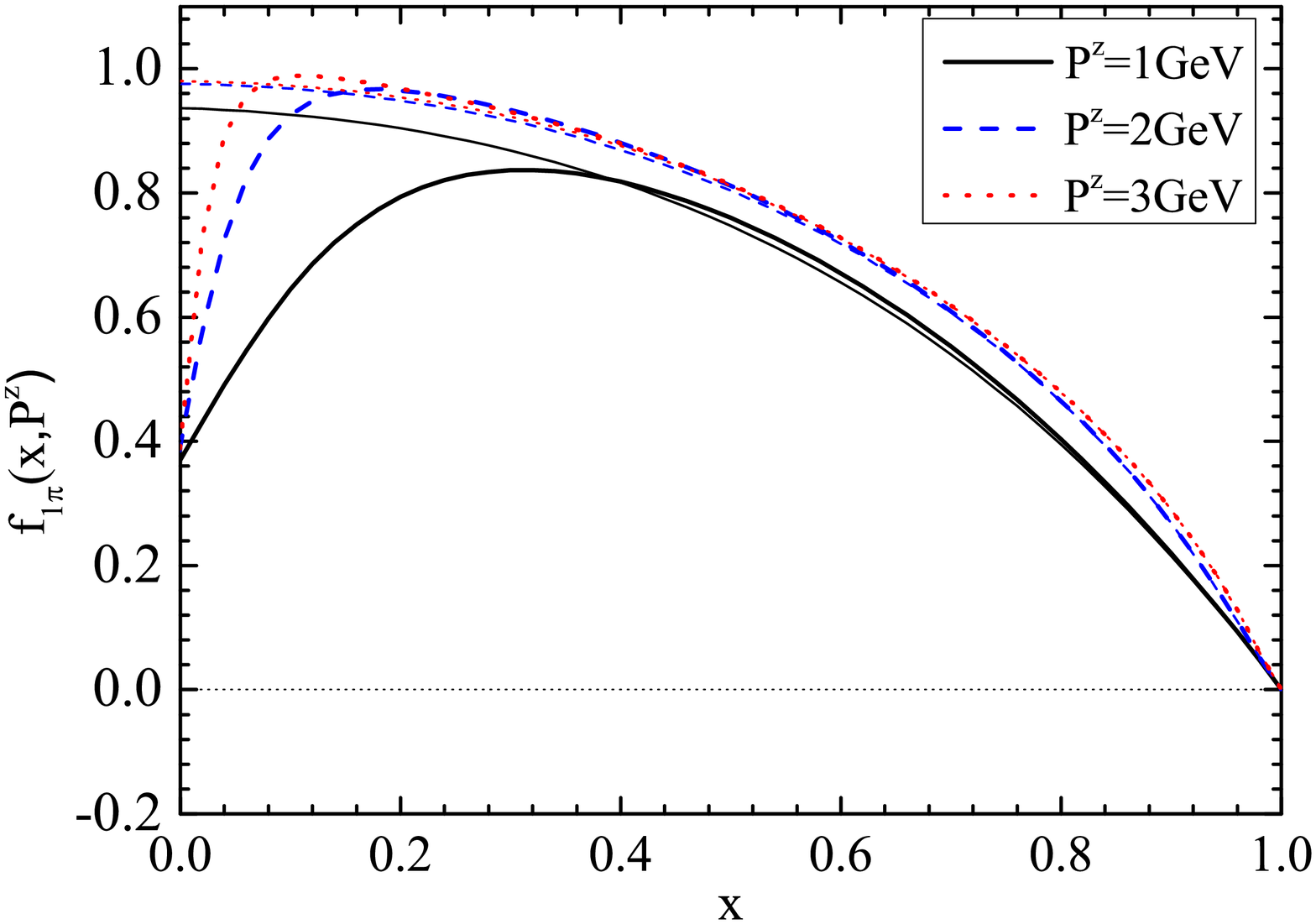}\\
  \caption{Left: quasi-PDF of the pion meson $f_{1\pi}(x; P^z)$ at $P^z= 1$ GeV (solid curves), 2 GeV (dashed curves) and $3$ GeV (dotted curves) respectively. The dashed-dotted line shows the standard pion PDF as a comparison. Right: comparison of the quasi-PDF from the contour integration approach (thick lines) and the quasi-PDF from the cut-diagram approach (thin lines). }\label{fig:qpdfx}
\end{figure*}

\begin{figure}
  \centering
  \includegraphics[width=0.49\columnwidth]{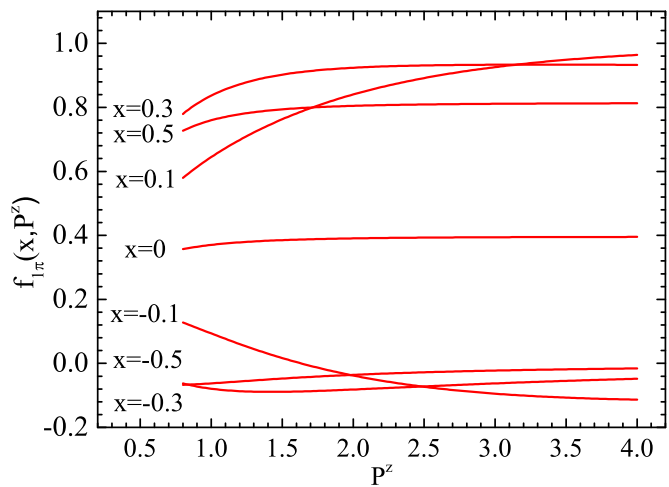}
  \caption{Quasi-PDF of the pion meson as function of $x$ for different $P^z$ values.}\label{fig:qpdfpz}
\end{figure}

In the left panel of Fig.~\ref{fig:qpdfx}, we plot the quasi-PDF of the pion meson in the spectator model as the function of $x$.
The solid line, dashed line and dotted line depict $f_{1\pi}(x;P^z)$ at $P^z=1 $ GeV, 2GeV, and 3 GeV, respectively.
The dashed-dotted line corresponds to the standard pion PDF  $f_{1\pi}(x)$ calculated from Eq.~(\ref{eq:pipdf}).
We find that in the region $x>0.4$, the quasi-PDF at $P^z=2$ or 3 GeV is very close to the standard PDF, both in shape and size.
This is different from the spectator result for the proton for which the quasi-PDF and the standard PDF can be very different at large $x$ region.
In the negative $x$ region, although the size of  $f_{1\pi}(x;P^z)$ is much smaller than that in the positive $x$ region, it is clearly nonzero.
In the positive $x$ region, the size of the quasi-PDF increases with increasing $P^z$.
In the region $x<-0.2$, the quasi-PDF is negative.
We have numerically verified that, when choosing very large values for $P^z$ ($P^z\gg 1$ GeV), $f_{1\pi}(x;P^z)$ approaches to $f_{1\pi}(x)$.
Also, an interesting feature in our model is that $f_{1\pi}(x;P^z)$ for different $P^z$ are almost same at $x=0$.

In the right panel of Fig.~\ref{fig:qpdfx}, we compare the quasi-PDF calculated from the contour integration (thick lines) and that calculated from the cut-diagram approach (thin lines) at different $P^z$ values.
We find that, in the large positive $x$ region, the two approaches are consistent.
However, in the small $x$ region, there is substantial difference between the two approaches.
It comes from the last four terms on the right hand side of Eq.~(\ref{eq:quasi-pdf}).
In the negative $x$ region, the quasi-PDF from the cut diagram approach vanishes, since the parton energy is required to be positive.

Fig.~\ref{fig:qpdfpz} shows the $P^z$-dependence of the quasi-PDF $f_{1\pi}(x;P^z)$.
It can be seen that there is almost no $P^z$ dependence at $x=0$.
This is consistent with the results in Fig.~\ref{fig:qpdfx}.
In the smaller $|x|$ region ($|x|=0.1$), one can see a strong $P^z$ dependence of $f_{1\pi}(x;P^z)$; while in the larger $|x|$ region ($|x|=0.5$), the $P^z$ dependence of $f_{1\pi}(x;P^z)$ is rougher small, and
the curves turn to flat at $P^z\sim 2$ GeV, which seems a scale reachable for current calculations of quasi-PDFs in lattice QCD~\cite{Chen:2018fwa}.

\begin{figure*}
  \centering
  \includegraphics[width=0.45\columnwidth]{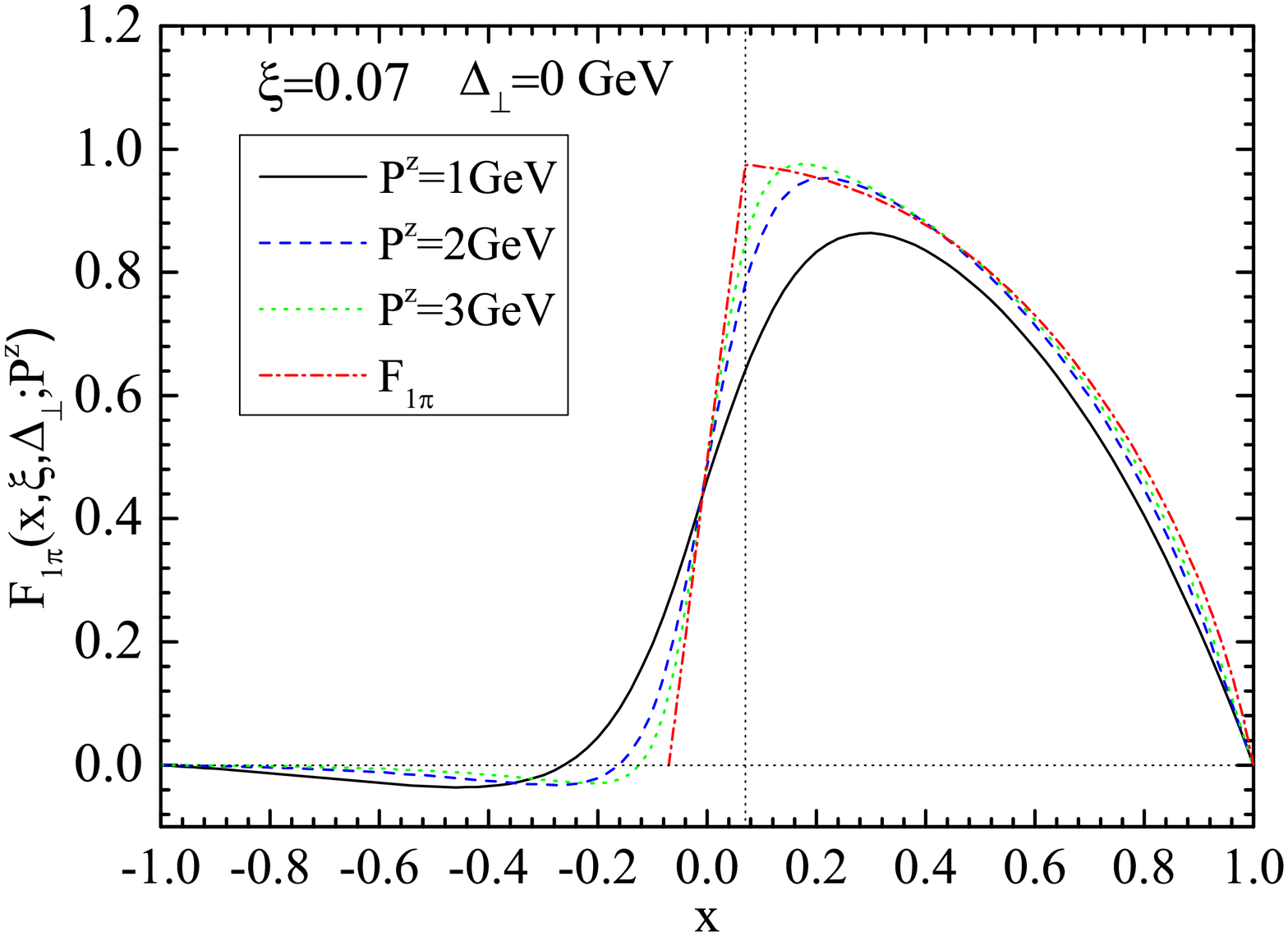}
  \includegraphics[width=0.45\columnwidth]{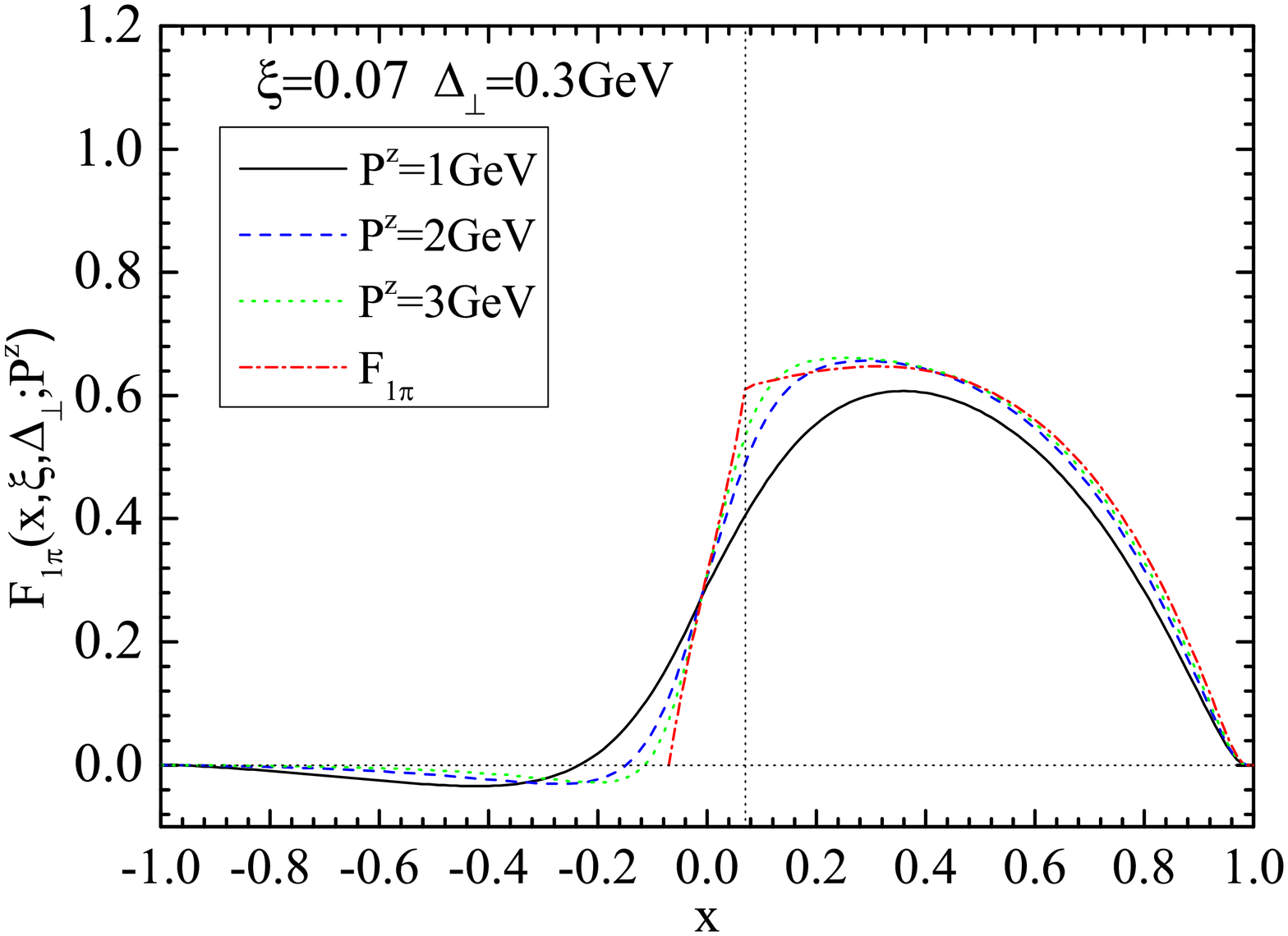}\\
  \includegraphics[width=0.45\columnwidth]{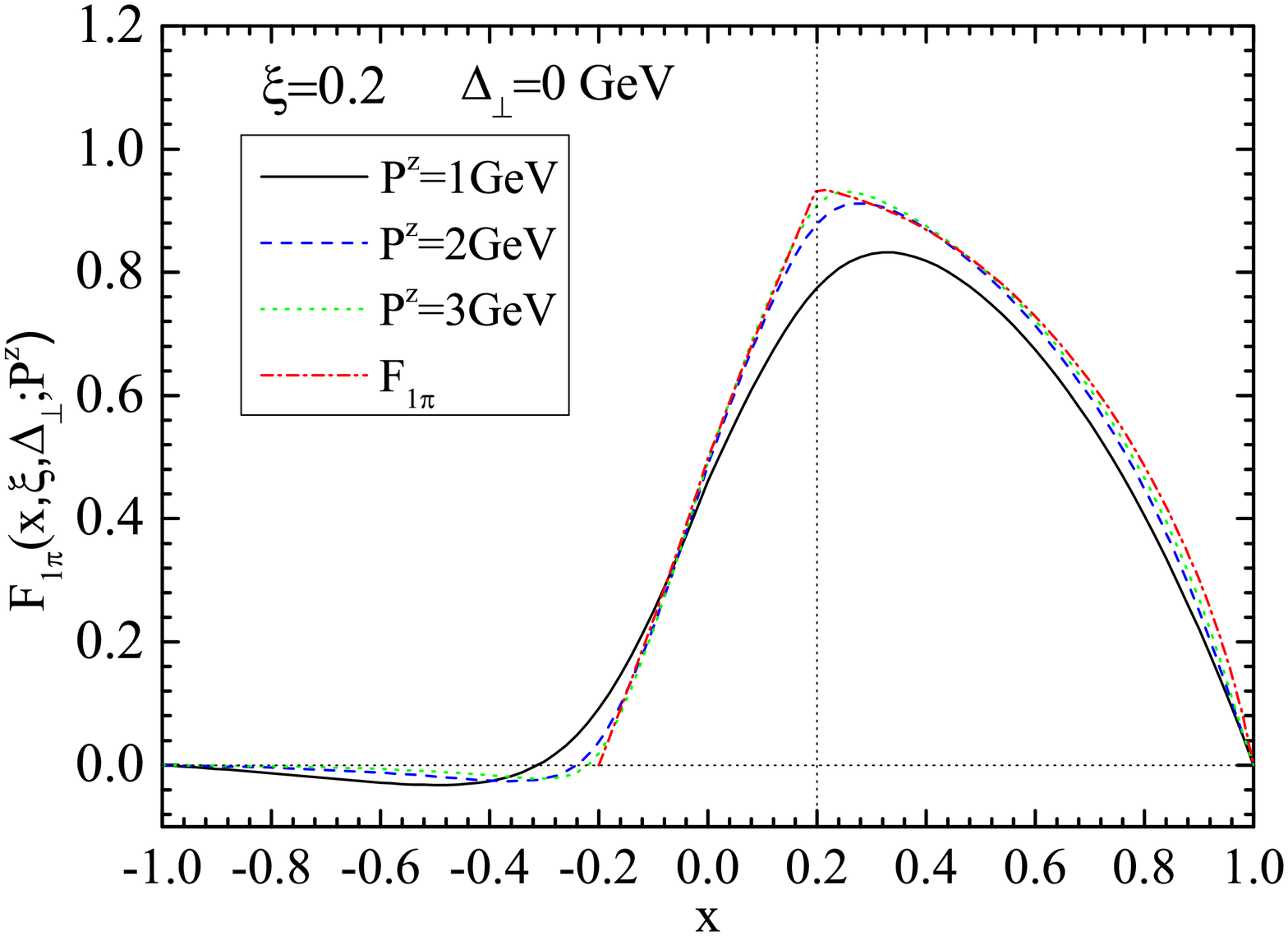}
  \includegraphics[width=0.45\columnwidth]{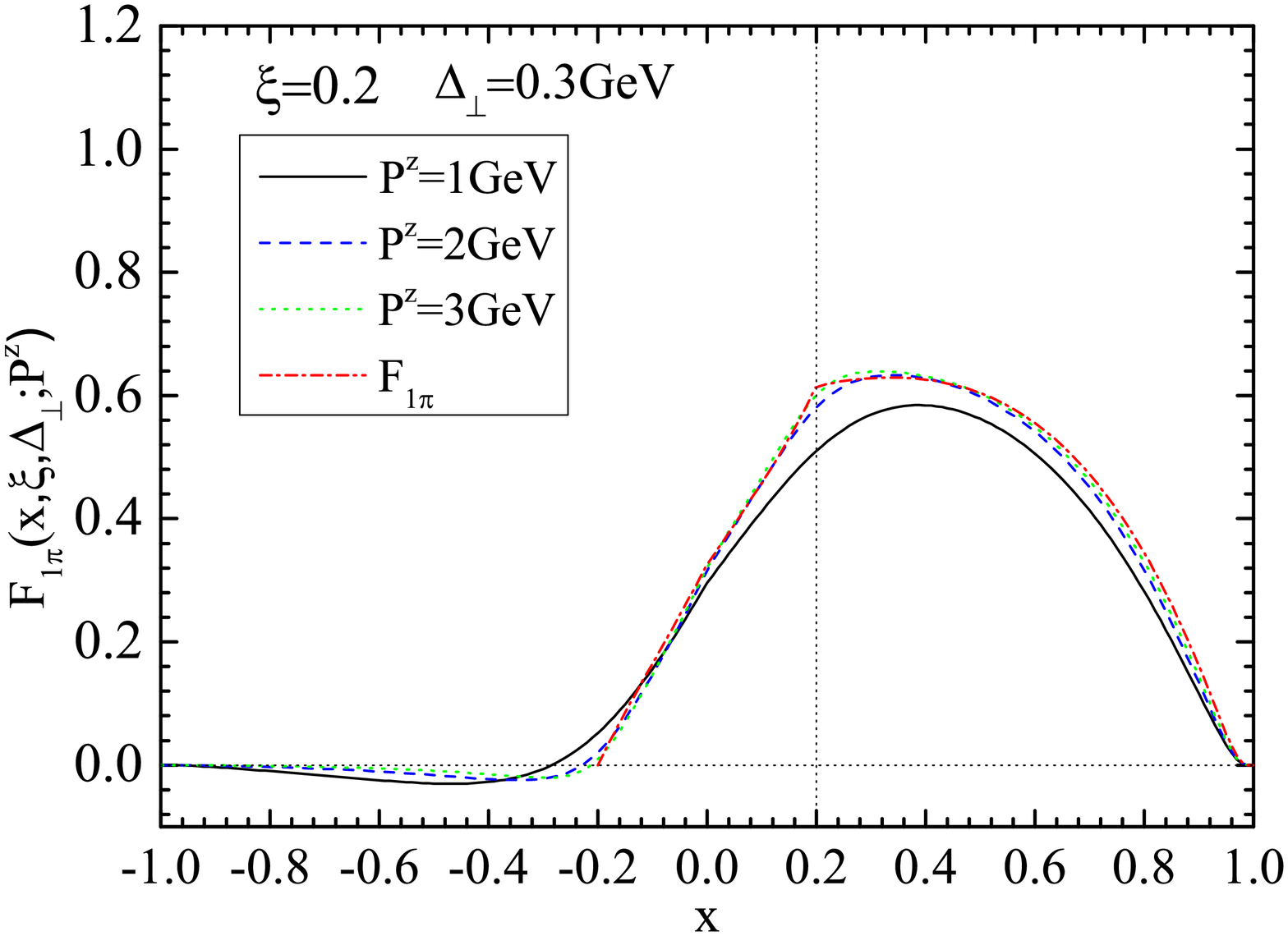}\\
  \includegraphics[width=0.45\columnwidth]{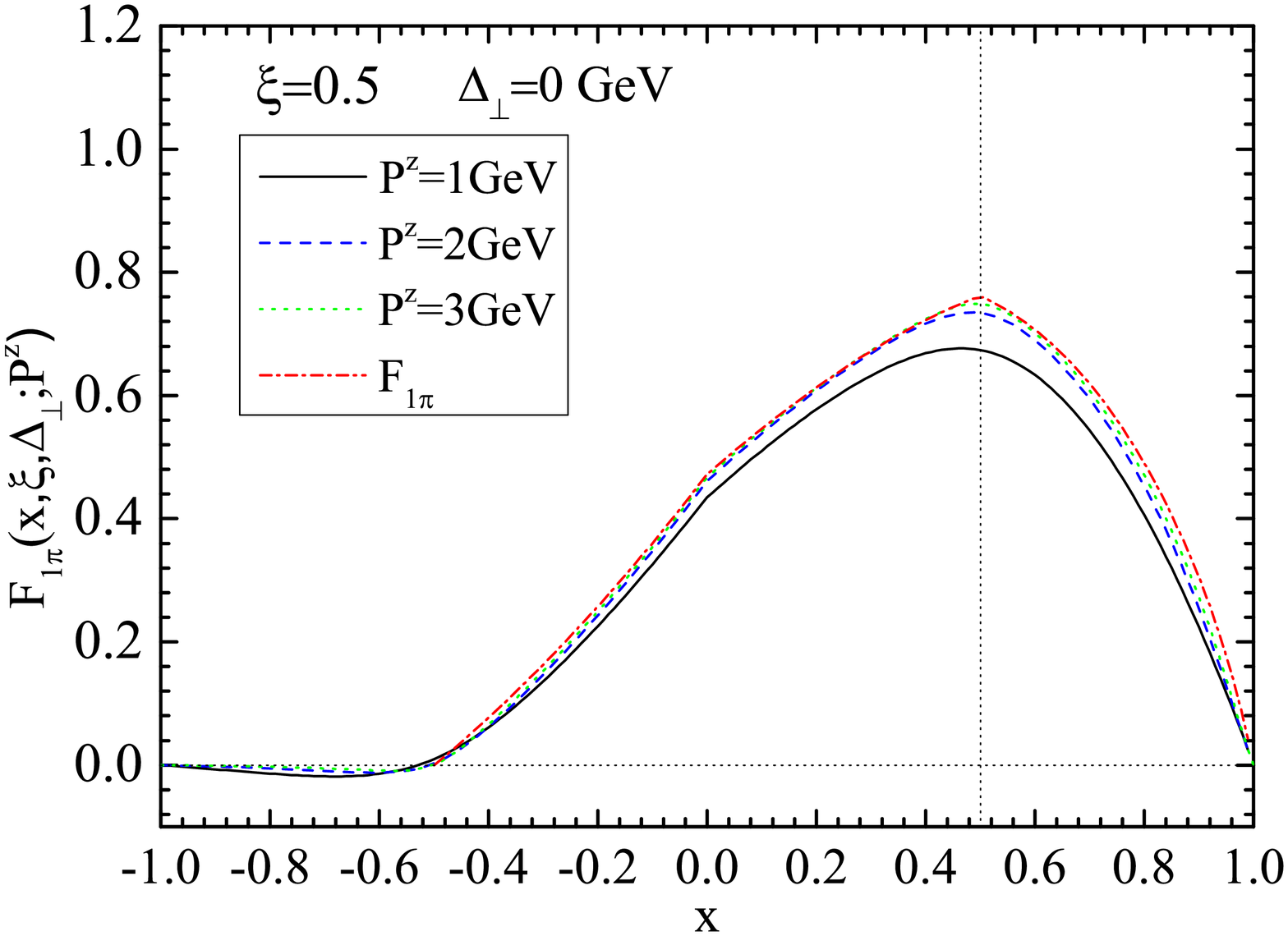}
  \includegraphics[width=0.45\columnwidth]{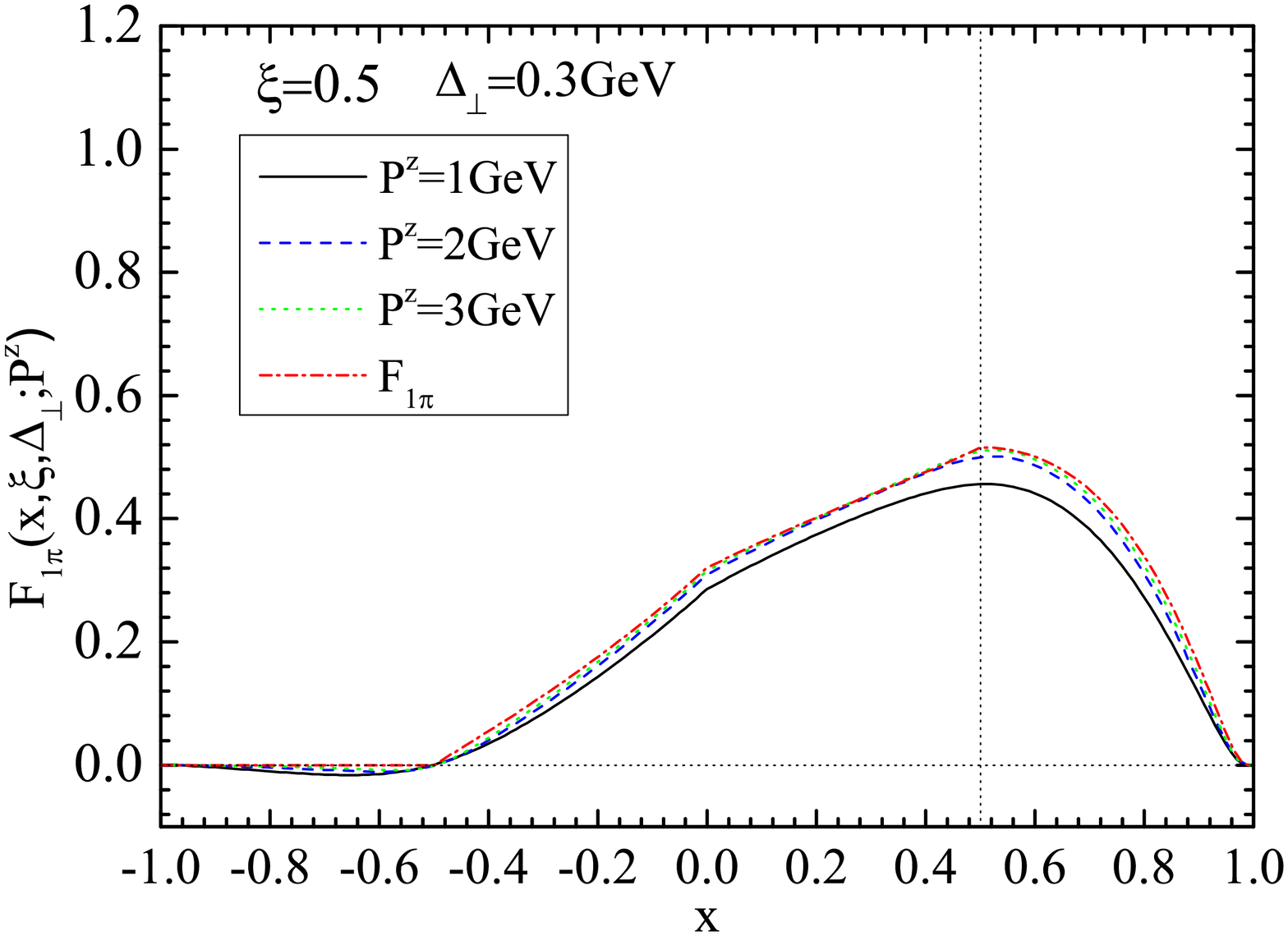}\\
  \caption{Quasi-GPD $F_{1\pi}(x,\xi,\Delta_\perp; P^z)$ as the function of $x$ at different $P^z$, $\xi$, and $\Delta_\perp$ values.
  The upper, central and right panels show $F_{1\pi}(x,\xi,\Delta_\perp; P^z)$ at $\xi= 0.07$, $0.2$ and $0.5$ respectively.
  The left and right panels depict $F_{1\pi}(x,\xi,\Delta_\perp; P^z)$  at $\Delta_\perp =0$ GeV and $\Delta=0.3$ GeV, respectively.
  The solid, dashed and dotted curves depict the quasi-GPD at $P^z= 1$ GeV, 2 GeV and $3$ GeV respectively.
  The dashed-dotted curves plot the standard GPD $F_{1\pi}(x,\xi,\Delta_\perp)$ for comparison.}\label{fig:quasigpd}
\end{figure*}

In Fig.~\ref{fig:quasigpd}, we plot the quasi-GPD $F_{1\pi}(x,\xi,\Delta_\perp;P^z)$ of the pion meson as the function of $x$ at different $P^z$, $\xi$, and $\Delta_\perp$ values.
The upper, central and right panels show $F_{1\pi}(x,\xi,\Delta_\perp;P^z)$ at $\xi= 0.07$, $0.2$ and $0.5$, respectively.
The left panel depicts $F_{1\pi}(x,\xi,\Delta_\perp; P^z)$ at $\Delta_\perp =0$ GeV, while the right panel shows $F_{1\pi}(x,\xi,\Delta_\perp;P^z)$ at $\Delta=0.3$ GeV.
The solid, dashed and dotted curves in each panel correspond to the quasi-GPD at $P^z= 1$ GeV, 2 GeV and $3$ GeV respectively.
The dashed-dotted curves plot the standard GPD $F_{1\pi}(x,\xi,\Delta_\perp)$ for comparison.
The quasi-GPD is continuous in the whole $x$ region. One can see that the standard GPD is also continuous, while its first-derivative is discontinuous at $x=\xi$.
We find that the quasi-GPD peaks in the region around $x\sim\xi$.
In the case $xi$ is small (such as $\xi=0.07$), one can see that the size of the quasi-GPD at low $P^z$ (=1 GeV) is smaller than the size of the standard GPD.
In the region $x<-\xi$, the quasi-GPD is nonzero, and mostly negative.
This is in contrast to the result of the standard GPD, which vanishes in the $x<-\xi$ region in the spectator model.
The quasi-GPD in the $x<-\xi$ region is much smaller than that in the region $x>\xi$.
In the large $\xi$ region, the quasi-GPD at different $P^z$ is very close to the standard GPD, namely, they have similar size and shape.
This result implies that the information on standard GPDs may be obtained from quasi-GPDs through nonperturbative methods such as lattice calculation~\cite{Chen:2019lcm}.
One can also find that, for small $\xi$ ($\xi<0.2$), the sizes of the quasi-GPD and the standard GPD decrease very quickly when $x$ decreases from $\xi$ to $-\xi$.
While in the large $xi$ region, a ``broadening" of the x-dependence of the quasi-GPD and the standard GPD can be observed.
For the $\Delta_\perp$-dependence of the pion quasi-GPD, generally the size of the quasi-GPD decrease with increasing $\Delta_\perp$.

\begin{figure*}
  \centering
  \includegraphics[width=0.49\columnwidth]{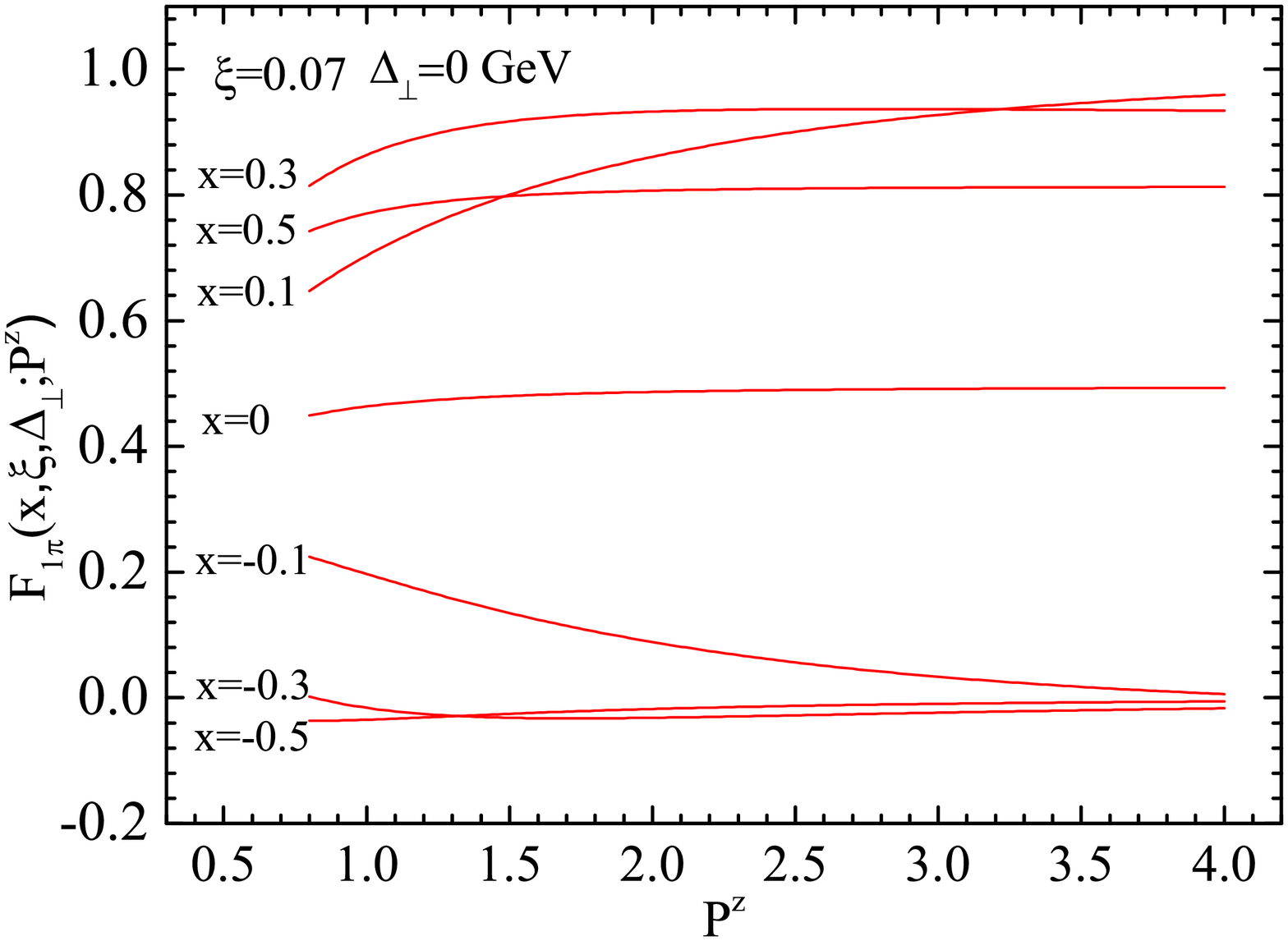}
    \includegraphics[width=0.49\columnwidth]{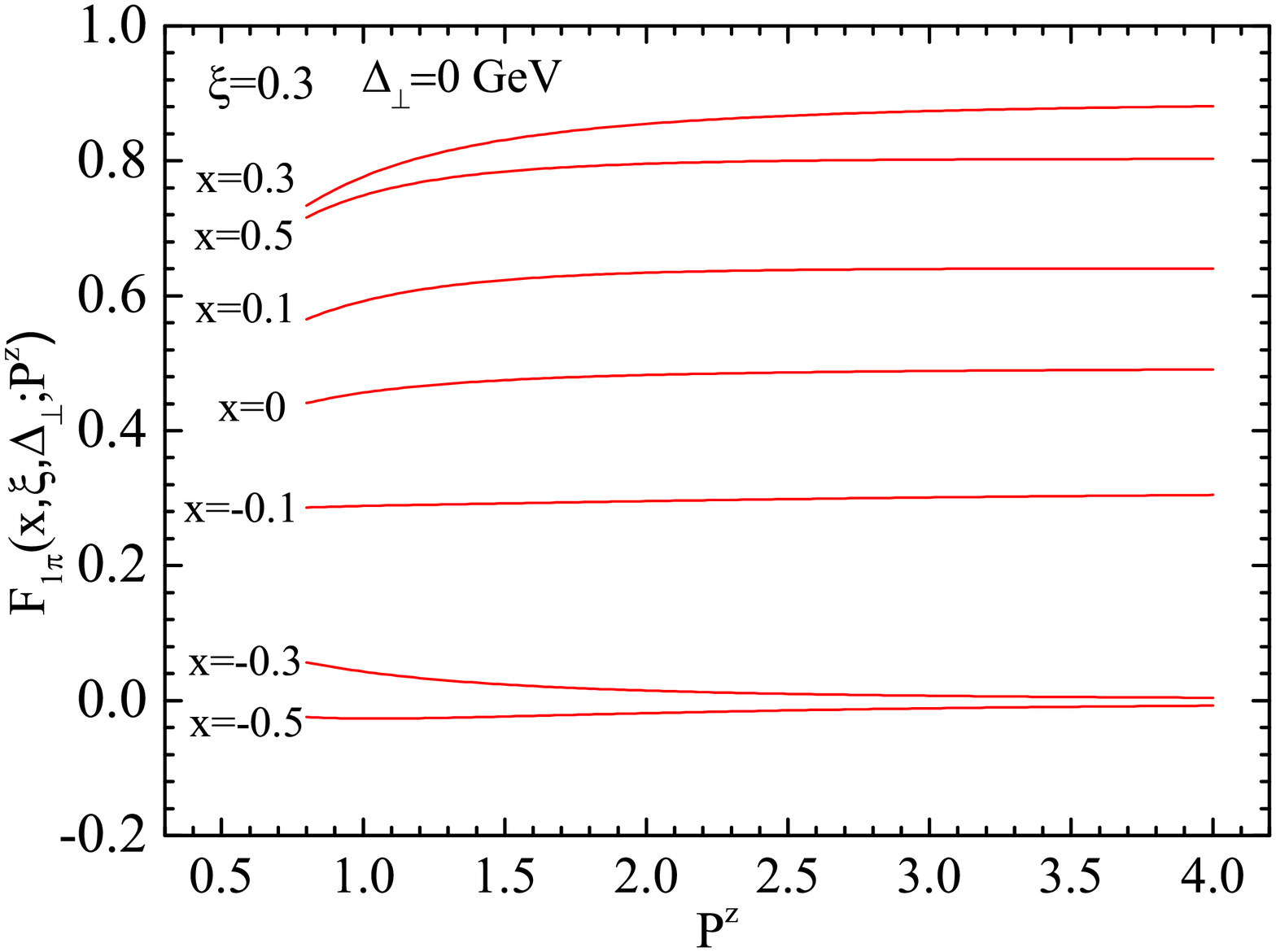}\\
  \caption{$P^z$ dependence of the quasi-GPD $F_1^e(x,\xi,\Delta_\perp; P^z)$ of the pion meson for different $x$ values at $\xi= 0.07$ (left panel) and $0.3$ (central panel), respectively.
  The solid, dashed and dotted curves depict the quasi-GPD at $P^z= 1$ GeV, 2 GeV and 3 GeV respectively.}\label{fig:qgpdppz}
\end{figure*}

To better illustrate the $P^z$ dependence of the Quasi-GPD, in Fig.~\ref{fig:qgpdppz}, we plot $F_{1\pi}(x,\xi,\Delta_\perp;P^z)$ as the function of $P^z$ at different $x$ values.
The left and right panels correspond to the results at $\xi=0.07$ and $\xi=0.3$, respectively. Without lose of generality, in both cases we choose $\Delta_\perp=0$.
We find that in the small $P^z$ region the $P^z$ dependence can change with the change of $\xi$.
In the large $P^z$ the quasi-GPD varies slowly.
In the case of $\xi=0.3$, we find that, even in the region $P^z$ is not too large (such as $P^z=2$ GeV),
the quasi-GPD almost becomes flat, which verifies the finding in Fig.~\ref{fig:quasigpd}.

\section{Summary}

In this paper, we have calculated the leading-twist quasi-PDF and the quasi-GPD of the pion meson using a spectator-antiquark model.
The quasi-PDF and quasi-GPD are defined via inserting the matrix $\gamma^3$ in the spatial quark-quark correlator.
In order to obtain the analytical results for these distributions, we have performed $k^0$-integral using the contour integration.
We have found that the quasi-PDF and quasi-GPD can reduce to the analytical results of the corresponding standard PDF and GPD in the limit  $P^z\rightarrow \infty$.
We have used an exponential form factor for the pion-quark-antiquark coupling to regularize the divergence in the high $k_T$ region, and we have focused on the kinematical region $-1\leq x \leq 1$, in which the analytical results of the quasi distributions are continuous.
We also calculated the quasi-PDF using the cut-diagram approach for comparison.
The parameters of the model have been determined by fitting the model result of the unpolarized PDF $f_{1\pi}(x)$ to the GRV parametrization to obtain more realistic results.
Our numerical calculation shows that, the quasi-PDF and the standard PDF of the pion meson almost coincide with each other in the region $x>0.2$.
In the region $P^z$ or $\xi$ is not small, the quasi-GPD of the pion meson is close to the standard GPD.
The $P^z$ dependence of the curve also shows that the quasi distributions in the intermediate and large $x$ region turns to flat when $P^z>2$ GeV.
In summary, our study have provided model implications and constraints on the quasi-PDF and quasi-GPD of the pion meson, which in general supports the idea of using quasi distributions to get information on standard distributions in lattice QCD.

\section*{Acknowledgements}

This work is supported in part by the National Natural Science Foundation of China (Grant Nos.11575043 and 11747086 ), and by the Young Backbone Teacher Training Program of Yunnan University. Z.~M. is supported by Yunnan Provincial New Academic Researcher Award for Doctoral Candidates.

\end{document}